\def\mypprint{\global\let\ifmypprint\iftrue}
\global\let\iftorefs\iffalse
\def\torefs{\global\let\iftorefs\iftrue}
\global\let\dofloatfig\iffalse
\def\floatthefig{\let\dofloatfig\iftrue}
\mypprint
\ifmypprint
\documentstyle[pre,aps,multicol,epsf,rotate,twoside]{revtex}

\draft
\floatthefig
\else
  \iftorefs
    \catcode`\@=11
    
    \def\figure{\let\@capwidth\columnwidth\@float{figure}}
    \let\endfigure\end@float
    \@namedef{figure*}{\let\@capwidth\textwidth\@dblfloat{figure}}
    \@namedef{endfigure*}{\end@dblfloat}
    \catcode`\@=12
     \floatthefig
  \fi
\fi
\input epsf.tex
\begin{document}

\twocolumn[\hsize\textwidth\columnwidth\hsize\csname @twocolumnfalse\endcsname
\title{Twirling and Whirling: Viscous Dynamics of Rotating 
Elastica}

\author{Charles W. Wolgemuth,$^{1}$ Thomas R. Powers,$^{3}$
and Raymond E. Goldstein$^{1,2}$}
\address{$^{1}$Department of Physics and $^{2}$Program in Applied Mathematics}
\address{University of Arizona, Tucson, AZ  85721}
\address{$^{3}$Division of Engineering and Applied Sciences, Harvard University, Cambridge, MA  02138}

\date{November 8, 1999}

\maketitle

\begin{abstract}
Motivated by diverse phenomena in cellular biophysics,  including
bacterial flagellar motion and DNA 
transcription and replication, we study the
overdamped nonlinear dynamics of a rotationally forced filament with
twist and bend elasticity.  Competition between twist injection,
twist diffusion, and writhing instabilities is 
described by a novel pair of coupled PDEs for twist and bend evolution. 
Analytical and numerical methods elucidate the twist/bend coupling and
reveal two dynamical regimes separated by a Hopf bifurcation:
(i) diffusion-dominated axial rotation, or {\it twirling}, and (ii) steady-state
crankshafting motion, or {\it whirling}.  The consequences of these
phenomena for self-propulsion are investigated, and experimental tests
proposed.

\end{abstract}

\vskip2pc]

Dynamics and stability of 
rotationally forced elastic filaments arise in several important biological 
settings involving
bend and twist elasticity at low 
Reynolds
number.  In the context of DNA replication, when two daughter strands 
are produced from
a duplex, it was
noted~\cite{levinthal} long ago
that energy dissipation for rotations about the filament axis 
is so much smaller than that for transverse motions that axial
``speedometer-cable" motions are favored, and are energetically and
topologically feasible.  During DNA 
transcription, in which a
polymerase protein moves down the double-stranded filament, 
progressive unwinding of the helix can lead to an
accumulation of local twist that may 
induce ``writhing" instabilities of the filament~\cite{wang}.  
Energetic and dynamical
aspects of these processes are of great current interest~\cite{marko98,phil}.

At the cellular level, bacteria are propelled through 
fluids by helical flagella turned by rotary motors 
in the cell wall~\cite{flag_reviews}. 
Recent studies~\cite{namba} have revealed the details of
two competing crystal structures assumed by 
flagellin, the protein
building block of flagella, corresponding to helices
of opposite chirality.
Both local and
distributed torques can change the conformation of
flagella; during swimming 
these motors episodically reverse direction~\cite{berg1}, and the resultant
torques can induce transformations between these 
states~\cite{chirality_flipping}, while uniform 
flow past a pinned flagellum may
induce such chirality inversions \cite{hotani}.

\begin{figure}
\epsfxsize=2.5truein
\centerline{\epsffile{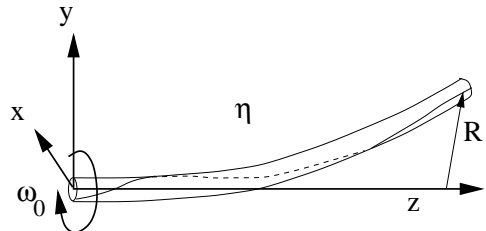}}
\smallskip
\caption{An elastic filament, rotated about $z$ at the left end, 
surrounded by a fluid of viscosity $\eta$. 
\label{fig1}}
\end{figure}

To elucidate fundamental processes
common to these systems, 
we consider here the model problem shown in 
Fig. \ref{fig1}: a slender elastic filament in
a fluid of viscosity $\eta$, rotated at one end at frequency $\omega_0$ 
with the 
other free.  We study 
competition between three processes: 
twist injection at the rotated end,
twist diffusion, and writhing.  Analytical and numerical
methods reveal two dynamical regimes of motion:  {\it twirling}, in
which the straight but twisted rod rotates about its centerline,
and {\it whirling}, in which the centerline of the rod writhes and
crankshafts around the rotation axis in a steady state.

This work is a natural outgrowth
of recent studies of forced elastica 
in the plane~\cite{ehd1,camalet},
and dynamic twist-bend coupling~\cite{kt,kamien,ehd3}.
The balance considered between elastic and viscous stresses
complements that between elasticity and inertia
in the inviscid limit 
(as in whirling shafts~\cite{love,whirling_shafts}), where
twist waves propagate~\cite{love,tabor}. 
 
An elastic filament is characterized by 
its radius $a$, 
contour length $L$, 
bending modulus $A$, and twist modulus $C$.  
The total elastic energy cost ${\cal E}$ for curvature $\kappa$ and 
twist density $\Omega$ is 
an integral over arclength $s$ parameterizing the position ${\bf r}(s,t)$ of 
the filament centerline~\cite{love}, 
\begin{equation}
{\cal E}=\int_0^L\!\!{\rm d}s\left({A\over2} \kappa^2 +{C\over2}\Omega^2 -\Lambda\right)~,
\label{energystrains}
\end{equation}
where the Lagrange multiplier $\Lambda$ 
enforces local inextensibility,
$({\bf r}_t)_s\cdot{\bf r}_s=0$.   
Thus arise two dimensionless ratios: 
$\Gamma\equiv C/A$,
and the aspect ratio $a/L$.
At zero Reynolds number, elastic forces per length
${\bf f}\equiv-\delta{\cal E}/\delta{\bf r}$~\cite{exactforce}
balance the viscous drag from 
slender-body hydrodynamics~\cite{keller}:
\begin{equation}
\zeta_{\|}\hat{\bf t}\hat{\bf t}\cdot{\bf r}_t
+\zeta_{\perp}({\bf I}-
\hat{\bf t}\hat{\bf t})\cdot{\bf r}_t={\bf f},
\label{resistive}
\end{equation}
where $\hat{\bf t}={\bf r}_s$ is the unit tangent,
and the transverse and longitudinal drag coefficients are 
$\zeta_{\perp}\simeq 2\zeta_{\|}\simeq 4\pi\eta/[\ln(L/2a)+c]$,
with $c$ a constant of order unity~\cite{keller}.
Likewise, the axial elastic torque per unit length 
$m=C\Omega_s$~\cite{ehd3,love} balances
the local rotational drag:  $m=\zeta_r\omega$,
where $\omega(s,t)$ is the local rotational velocity about $\hat{\bf t}$ 
and  
\dofloatfig
\fi 

\noindent  $\zeta_r\simeq 4\pi\eta a^2$~\cite{keller}. 
We also define 
$\epsilon^2\equiv \zeta_r/ (\zeta_{\perp}L^2)$, so $\epsilon\sim (a/L)$, 
apart from logarithmic corrections. 

The dynamics are closed by a geometric constraint,
\begin{equation}
\Omega_t=\omega_s+(-\Omega{\bf r}_s+{\bf r}_s\times{\bf r}_{ss})\cdot
[{\bf r}_t]_s~,
\label{twistdot}
\end{equation}
which shows how twist changes due to differential rates of angular
rotation, stretching ({\it e.g.}, extension of a straight, twisted rod decreases
$\Omega$), and out-of-plane bending motions 
(writhing)~\cite{kt,kamien,ehd3}. 
The constraint (\ref{twistdot})
is a conservation law for twist density 
$\Omega$, with twist current $-\omega$, and with the stretching and 
writhing terms acting as sinks or sources.
The velocities $\omega$ and
${\bf r}_t$ enter (\ref{twistdot}) through their space derivatives,
since rigid motions cannot change $\Omega$.
With the local torque balance $\zeta_r\omega = C\Omega_s$
and assuming inextensibility we obtain
\begin{equation}
\Omega_t = {C\over\zeta_r}\Omega_{ss} + {1\over{\zeta_{\perp}}}
{\bf r}_{s}\times{\bf r}_{ss}\cdot{\bf f}_s~.
\label{formal_dynamics}
\end{equation}    
The second term of (\ref{formal_dynamics})
is nonzero~\cite{exactforce1} 
when the filament is both out of elastic equilibrium (${\bf f}\ne 0$) 
and either nonplanar (with torsion $\tau\ne 0$) or twisted ($\Omega \ne 0$), 
and then acts as a sink or source of twist.

For boundary conditions, we assume the forced end (${\bf r}(0)= 0$) of the 
rod is clamped
(${\bf r}_s(0)={\bf{\hat{z}}}$) and the free end experiences no force or torque
(${\bf r}_{ss}(L)={\bf r}_{sss}(L)=0; \Omega(L)=0$).  Local torque balance
sets $\Omega_s(0)=\zeta_r\omega_0/C$.

Before solving these PDE's numerically, we use dimensional analysis
to understand the main features of the motion.
We focus on
small-amplitude bend and twist deformations of a straight filament 
(thus ignoring
$\zeta_{\|}$).  Of the seven parameters 
$(A,C,L,a,\zeta_r,\zeta_{\perp},\omega_0$), four remain after introducing
$\Gamma$ and $\epsilon$ and noting that $a$ only appears through $\zeta_r$, 
so suitable rescalings of length, time, and 
$\Omega$ 
leave only one control parameter.  This can be chosen proportional to the
rotation frequency $\omega_0$.  

For low turning rates $\omega_0$, the filament remains straight,
with twist diffusion and injection balancing.
The twist density at the clamped end follows from a balance of
viscous and elastic twisting torques, $\zeta_r\omega_0 L\sim C\Omega$, or
\begin{equation}
\Omega(0) \sim {\zeta_r\omega_0 L\over C}\equiv\Omega_0~.
\label{omega_zero}
\end{equation}
Instability occurs when the
the twist torque $C\Omega$ is comparable to
the filament buckling torque $A/L$~\cite{love}.
At this point, the balance of viscous and twist torques (\ref{omega_zero})
implies
\begin{equation}
\omega_c\sim {A\over \zeta_r L^2}\sim \left({a\over L}\right)^2{E\over \eta}
\sim {k_BT\over \zeta_r L_p} \left({L_p\over L}\right)^2~,
\label{omega_crit}
\end{equation}
where the second and third result follow from the 
relations $A=(\pi/4) a^4 E$ between the bending modulus 
and the Young's modulus $E$~\cite{love} and $A=k_BTL_p$, with
$L_p$ the persistence length.  
Interestingly, 
$\omega_c$ is independent of the twist modulus $C$, and
since the twist density scales with the drag,
$\omega_c$ varies inversely with $\zeta_r$. 

The result (\ref{omega_crit}) is central, for naive dimensional analysis
predicts $\omega_c\sim E/\eta$. With $E \sim 10^7 \mbox{dynes}/\mbox{cm}^2$
and $\eta\sim 0.01$ Poise as for rubber in 
water,
$\omega_c$ would be enormous if not for
the prefactor $(a/L)^2$.  Since $a/L$ is reasonably $10^{-3}$, we 
find $\omega_c\sim 10^3$ s$^{-1}$, similar to
flagella rotation
rates~\cite{flag_reviews} and
achievable in the laboratory.
The rightmost form in (\ref{omega_crit}) shows readily
the frequency scales for systems of varying length and stiffness.  Consider
the elastica DNA ($L_p\sim 5\times 10^{-6}$~cm, $a\sim 10^{-7}$~cm), 
microtubules ($L_p\sim 0.5$~cm, $a\sim 10^{-6}$~cm), and 
bacterial filaments ($L_p\sim 25$~cm, $a\sim 3\times 10^{-5}$~cm) 
\cite{thwaites}.  The frequencies $k_BT/\zeta_r L_p$ are then $8\times 10^{6}$ s$^{-1}$, 
$0.8$ s$^{-1}$, and $1.3\times 10^{-5}$ s$^{-1}$, respectively.  Thus, for the
instability to appear at, say, $10^{3}$ s$^{-1}$ requires 
a minimum ratio $L/L_p$ of $90$, $0.03$, and $10^{-4}$.  A strand of
DNA with $L/L_p \sim 10^2$ is clearly not straight in isolation, so
this instability would be hard to realize in DNA, but
the stiffer examples of microtubules and bacterial filaments are indeed
candidates.

Linearizing (\ref{resistive}) and (\ref{formal_dynamics}) about a straight
filament along $\hat{\bf z}$, with ${\bf r}\approx s\hat{\bf z}+{\bf r}_\perp$,
we see that twist diffuses with diffusion constant
$C/\zeta_r$, while the backbone obeys  
a ``hyperdiffusion'' equation
$\zeta_{\perp}{\bf r}_{\perp t}= -A{\bf r}_{\perp ssss}+C\hat{\bf z}\times
(\Omega{\bf r}_{\perp ss})_s$. 
For crankshafting motions, we set 
${\bf r}_{\perp t}=\chi\hat{\bf z}\times{\bf r}_\perp$.
Thus we find two characteristic lengths~\cite{ehd1}, 
\begin{equation}
\ell_{\perp}(\chi)=\left(A/ \zeta_{\perp}\chi\right)^{1/4}~~ {\rm and}~~~
\ell_r(\omega_0)=\left(C/\zeta_r\omega_0\right)^{1/2}~.
\label{ellperp}
\end{equation}
These are analogous to
the penetration depth in the familiar theory of oscillations in
a viscous fluid~\cite{stokes}. 
The primary instability is given by $\ell_{r}(\omega)\sim L$.

The crankshafting frequency $\chi$ for whirling
can be estimated by 
assuming that the transverse drag, 
$\zeta_\perp\chi |{\bf r}_\perp|$,
is roughly equal to the elastic force per length, 
$A|{\bf r}_{\perp4s}|\sim A|{\bf r}_\perp|/L^4$.
Thus $\chi\sim A/\zeta_\perp L^4 \sim (a/L)^2\omega_c$ ($C\approx A$ for
typical materials~\cite{love}), and $\ell_{\perp}(\chi)\sim
\ell_{r}(\omega_c)\sim L$ at the transition.
The whirling rod 
does not undergo simple rigid body rotation; the 
speedometer-cable rotational motion is 
faster than the backbone crankshafting motion by a factor of $(L/a)^2$.
This steady-state shape is possible because diffusion can
homogenize the twist as fast as backbone motion
relieves it.
We describe this 
process quantitatively by 
integrating (\ref{twistdot})
along the rod for inextensible, steady-state ($\Omega_t=0$)
crankshafting.  
The difference $\Delta\omega$ in rotational velocities about
the local tangents at $s=L$ and $s=0$ is
\begin{equation}
-\Delta\omega=\chi\bigl[1-\hat{\bf z}\cdot\hat {\bf t}(L)\bigr].
\label{twistsink}
\end{equation}
Equivalently, $\Delta \omega$ is the injected twist
current minus the twist current leaving the free end.  Thus, 
writhing acts as a twist sink in steady-state crankshafting
motion when the rod's free end is not aligned with the $z$-axis.

When the twist diffusion time, $\zeta_r L^2/C$, 
is longer than the bending time $\chi^{-1}$,
buckling can relieve twist faster than it is replenished
by diffusion, and steady-state crankshafting 
would likely be unstable. One possible new behavior would consist of
repeated sequences of transient whirling followed by quiescence as
twist builds up anew.
Scaling arguments yield a critical frequency of
$\omega_0 \sim E/\eta$,
a factor of $\epsilon^{-2}$ higher than the rate at onset of the first
instability, and thus unreachable for typical materials. 

The bend relaxation time suggests the rescaling,
\begin{equation}
{\tilde t} \equiv \left(A/\zeta_{\perp} L^4\right)~.
\label{alpha}
\end{equation}
A natural pair of further rescalings of (\ref{formal_dynamics}) is
$\tilde s=s/L$, 
and $\tilde\Omega=\Omega L$.
If we parameterize the filament centerline 
(Fig. \ref{fig1}) as 
${\bf r}(s,t) = (X(s,t), Y(s,t), s-\delta(s,t))$, 
introduce the complex transverse displacement
$\xi = (X + iY)/L$, and expand the dynamics up to
third order in $\xi$ (immediately dropping the tildes), we obtain 
\begin{eqnarray}
\xi_t&=& -\xi_{4s}
 - \left(\Lambda\xi_{s}\right)_s
+ i\Gamma\left[\Omega\left(\xi_{ss}(1-\delta_s\right) 
+ \delta_{ss}\xi_s)\right]_s\nonumber\\
&&- \left[{1\over2}\left(\xi_{4s}^*\xi_s
+\xi_{4s}\xi_s^*\right)+\Lambda_s\right]\xi_s\nonumber\\
\Omega_t &=& {\Gamma\over\epsilon^2}\Omega_{ss} + {1\over2}\bigl[
i\left(\xi_{5s}^*\xi_{ss}-\xi_{5s}\xi_{ss}^*\right)\nonumber\\
&&+\Gamma\left(\left(\Omega\xi_{ss}^*\right)_{ss}\xi_{ss} 
+ \left(\Omega\xi_{ss}\right)_{ss}\xi_{ss}^*\right)\bigr] ~,
\label{weaknon}
\end{eqnarray}
where the inextensibility 
constraint $(1-\zeta_{\|}/\zeta_{\perp})({\bf r}_{ss}\cdot{\bf f})
+({\bf r}_s\cdot{\bf f}_s) \simeq {1\over2}({\bf r}_{ss}\cdot{\bf f})
+ ({\bf r}_s\cdot{\bf f}_s) = 0$ 
is expanded to set $\Lambda$ as
\[
\left(\Lambda - {3\over2}|\xi_{ss}|^2\right)_{ss}
= {1\over2} \left({\cal R}e(\xi_{4s}\xi^*_{ss})
+i\Gamma\Omega{\cal I}m(\xi_{3s}\xi_{ss}^*)\right).
\]
This constraint
also fixes $\delta_s\simeq {1\over2}\vert\xi_s\vert^2$.
As anticipated, apart from  
material properties $\Gamma$ and $\epsilon$,
the coupled twist/bend dynamics are governed by a single control
parameter $\omega_0$, which appears only in the (rescaled) boundary condition,
$\Omega_s(0)=\epsilon^2\omega_0/\Gamma=\alpha$.  The PDEs (\ref{weaknon})
are like those of 
excitable media \cite{hohenberg}, with a separation of
time scales derived from the aspect ratio $\epsilon$.  For the usual case
$\epsilon \ll 1$, twist is the fast variable and bends are slow.

{\it{Linearized Dynamics}:}
The twist profile in the straight filament ($\xi=0$)
satisfies $\Omega_t = (\Gamma/\epsilon^2)\Omega_{ss}$.  After 
transients die out, the steady-state profile is linear in $s$~\cite{marko98},
$\Omega =(\zeta_r\omega_0/ C)\left(s - L\right)~.$
Using this in the linearized filament evolution 
and taking
$\xi(s,t)=\xi(s)\exp(i\chi t)$, a rigidly rotating, neutrally stable shape,
we obtain
\begin{equation}
\chi\xi = i\xi_{4s} + \Gamma\alpha[(s-1)\xi_{ss}]_s~.
\label{linxieq}
\end{equation}

Numerical solution~\cite{numerical_recipes} of (\ref{linxieq}) 
yields a critical value 
$\alpha_c\simeq 8.9/\Gamma$ (confirming dimensional
analysis
of Eq. (\ref{alpha})), below which the rod is straight and executes 
only axial
 rotation (``twirling"), and above which the rod 
buckles and rotates (``whirls") at a frequency 
which for $\alpha \simeq \alpha_c$ is 
$\chi \approx 2.32 \Gamma\alpha$. This motion is the dynamical equivalent of the static writhing instability of a
twisted elastica~\cite{love}.  Inserting all numerical factors,
the critical frequency and
rotation rate (at onset) are
\begin{equation}
\omega_{c} \simeq  0.563 \left({a\over L}\right)^2{E\over \eta}~, \ \ \ \ 
\chi_c \simeq 20.9\left({a\over L}\right)^2\omega_{c}~.
\label{omega_crit1}
\end{equation}

{\it{Weakly Nonlinear Theory}:}
Numerical solution of (\ref{weaknon})
with a pseudospectral
method~\cite{ehd3,GnL}
shows that there is indeed a steady state beyond the bifurcation.
As $\alpha-\alpha_c$ increases, the shape becomes more helical.
The free end of the whirling filament experiences more drag than
\dofloatfig
\begin{figure}
\epsfxsize=2.1truein
\centerline{\epsffile{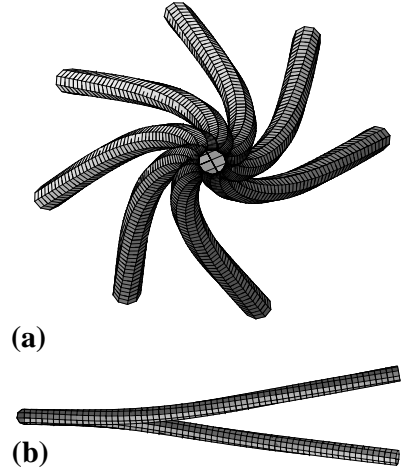}}
\smallskip
\caption{(a) Stroboscopic montage of the ``whirling'' filament,
viewed from along the $z$-axis, as it
rotates clockwise; (b) side view at two times, drawn to a different 
scale. 
\label{fig2}}
\end{figure}
\fi
\noindent
points
closer to the driven end and thus lags behind (Fig.~\ref{fig2}).  
Since $\Omega$ depends
quadratically on the backbone shape (see Eq. \ref{twistdot}), 
near $\omega_c$, where $\xi$ is small, the twist
density remains nearly linear in $s$.
Numerical studies show that the free end traces out a circle
with radius $R\sim (\omega_0-\omega_c)^{1/2}$:
a supercritical Hopf bifurcation~\cite{hohenberg}.  
This can be understood from
(\ref{twistsink}) and dimensional arguments for the displacement;
$\bigl[1-\hat{\bf z}\cdot\hat {\bf t}(L)\bigr] \simeq (R/L)^2$, so
$\omega(0) - \omega(L) \sim \chi (R/L)^2$.
The twist that can be 
relieved by diffusion is limited, so $\omega(L) \approx \omega_c$.  From the 
linear dynamics, $\chi \sim
\omega_c$, leading to $R \sim (\omega_0 - \omega_c)^{1/2}$.
\dofloatfig
\begin{figure}
\epsfysize=3.2truein
\centerline{\epsffile{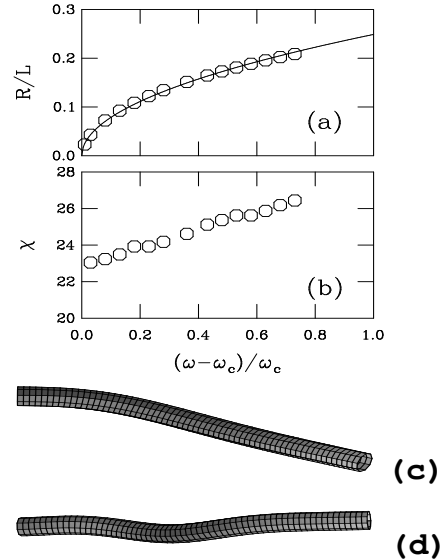}}
\bigskip
\caption{(a) Amplitude $R$ (with a square root fit) and
(b) crankshafting frequency of filament tip motion 
as a function of frequency offset from primary instability.  
(c) and (d); filament shapes for
$(\omega_0-\omega_c)/\omega_c = 0.27$ and $3.52$, with $\Gamma = 1$. 
\label{fig3}}
\end{figure}
\fi 

\dofloatfig
\begin{figure}
\epsfysize=2.2truein
\centerline{\epsffile{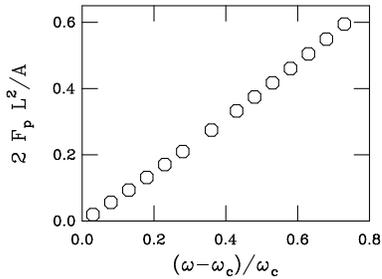}}
\bigskip
\caption{Propulsive force generated by steady-state whirling motions,
as a function of driving frequency. 
\label{fig4}}
\end{figure}
\fi

{\it{Swimming}:}
Chirality of the whirling filament breaks time-reversal
invariance of the motion, thereby allowing \cite{taylor} a net propulsive
force $F_p$ along $\hat {\bf z}$ to be generated.
The elastic propulsive force density is a total derivative, so the total force
is expressible in terms of the filament properties at its ends.
For the clamped/free boundary conditions used here, 
$F_p = -{\bf r}_{3s}(0)\cdot{\bf{\hat{z}}} - \Lambda(0) = {\bf r}_{ss}(0)
\cdot{\bf r}_{ss}(0) - \Lambda(0)$.  As shown in Fig. \ref{fig4}, $F_p$
rises linearly from zero near the bifurcation, as it is quadratic in 
the transverse displacement, which in turn has the supercritical
form shown in Fig. \ref{fig3}(a).
While we know of no organism that utilizes this precise mechanism for
self-propulsion, there is evidence 
in certain laboratory experiments \cite{walking}
for self-propulsion associated with
twist-induced whirling in growing bacterial
macrofibers constrained at one end.
Experiments are underway to explore further this possible connection.

The possibility of observing instabilities driven by
twist accumulation along an elastica hinges upon a balance of
material properties, fluid viscosity, and adequate forcing.
Flow- and rotation-induced bacterial 
flagellar conformational 
transitions~\cite{berg1,chirality_flipping,hotani} provide
proof-of-principle that this balance can be achieved {\it in vivo}.
Like flagella, fibers of {\it B. subtilis} cells
have adequate material properties (e.g. Young's modulus~\cite{thwaites})
and aspect ratio to display instabilities like those described here.
More complex phenomena are associated with instabilities of rotating helical
flagella, as described elsewhere~\cite{next_paper}.

We thank P. Nelson and C. Wiggins 
for discussions motivating this
work, and D. Coombs, A. Goriely, G. Huber, J.O. Kessler, S. Koehler,  
H.A. Stone, and especially N.H. Mendelson and J.E. Sarlls for ongoing 
collaborations.  
This work was supported by
NSF Grant DMR9812526 (REG); TRP thanks H. Stone for
support through the Harvard MRSEC  and
the Army Research Office Grant DAA655-97-1-014.

\end{document}
\end

\end{document}
\end